\documentclass[preprint,10pt]{elsarticle}

\usepackage{amssymb}
\usepackage{float}
\usepackage{amsmath}
\usepackage{setspace}
\usepackage[labelfont=bf]{caption} 
\usepackage{graphicx}
\usepackage{xcolor}
\usepackage{svg}
\usepackage{ulem}
\journal{Journal of Electronic Materials}
\begin{document}
\begin{frontmatter}

\title{Impact of Switching Layer Architecture on Power Consumption in RRAM}

\author[label1,label2]{John F. Hardy II\corref{cor1}}
\author[label2]{Jack Garrard}
\author[label2]{Guilherme S.Y. Giradini}
\author[label1,label2]{Carlo R. daCunha}

\affiliation[label1]{organization={Center for Materials Interfaces in Research and Applications, Northern Arizona University},
            addressline={700 Osborn Dr.}, 
            city={Flagstaff},
            postcode={86011}, 
            state={AZ},
            country={USA}}
\cortext[cor1]{Corresponding author. Email: jfh67@nau.edu}
\affiliation[label2]{organization={School of Informatics, Computer, and Cyber Systems, Northern Arizona University},addressline={1295 Knoles Drive},city={Flagstaff},postcode={86011},state={AZ},country={USA}}

\begin{abstract}
This work demonstrates that porous helical WO$_x$ architectures enable a distinct low-power regime for planar ITO/WO$_x$/ITO resistive random-access devices. While thin film and helical devices behave similarly at a 5 mA compliance, only helical devices sustain reproducible operation at $500$~$\mu$A, where RESET voltages reduce by $\mathord{\mathord{\sim}}60\%$, switching currents decrease by $68-75\%$, and SET/RESET power drops by $\mathord{\mathord{\sim}}89\%$ and $\mathord{\mathord{\sim}}83\%$. With helical devices operating at $500$~$\mu$A, the memory window expands $400$--$600\%$ due to selective suppression of high-resistive-state leakage, yielding both lower-power and improved read margin in a regime inaccessible to thin film devices. These results highlight geometry-driven field enhancement and confinement as practical design principles for low-power, high-margin resistive memories and point toward opportunities in transparent, flexible, and high-surface-area material systems.
\end{abstract}


\end{frontmatter}
\doublespacing

\section{Introduction}
Traditionally, silicon-based Flash memories are the current prevailing non-volatile storage devices, yet their fundamental operating principles present increasing limitations as device dimensions demand improved scalability and applications demand intensified operations \cite{Memories}. Flash memory encodes information by storing electrons in a floating gate within a metal-oxide-semiconductor field-effect transistor (MOSFET) \cite{Flash}. While Flash remains widely adopted due to its non-volatility, data retention capabilities, and relatively high storage density, it suffers from several critical limitations. These limitations consist of high program and erase voltages, long write/erase times, and limited endurance. Down-scaling below the $20$ nm node presents more challenges, such as charge leakage, increased variability, and degradation in reliability and yield  \cite{Flash_Cycles, Flash_Volts, FLASH_Prob}. More broadly, conventional memory technologies are constrained by the Von Neumann architecture, where physically separated memory and processing units lead to the well-known memory bottleneck \cite{Von_Bottle}. The limitations of conventional memory technologies are particularly detrimental in emerging applications such as edge computing, artificial intelligence, and real-time data processing, which necessitate high-speed and low-power memory solutions \cite{Von_NeuProb}. To overcome these constraints, extensive research has been directed to the development of emerging non-volatile memory technologies that simultaneously address performance and scalability requirements \cite{Mem_Bottle}. 

Among these, resistive random-access memory (RRAM) has received considerable attention for applications in both high-density memory and neuromorphic computing hardware \cite{RRAM_Bias}. RRAM is a two-terminal, non-volatile device in which data storage is achieved by modulating the resistance state through the application of an electric field \cite{RRAM_Overview}. Compared to conventional Flash memory, RRAM exhibits distinct advantages, including low operating voltage, fast switching speed, scalability to sub-$10$ nm technology nodes, and compatibility with three-dimensional integration \cite{RRAM1, RRAM2, RRAM3}. A typical RRAM device's architecture consists of a metal-insulator-metal thin-film (TF) stack, which offers both design simplicity and scalability \cite{RRAM_Compact}. In a vertical stack, the bottom electrode (BE) is typically grounded while a bias is applied to the top electrode (TE) to program the device by altering the resistance state of the insulating-layer, also referred to as the active-layer \cite{RRAM_Bias1}. In addition to the conventional vertical configuration, planar RRAM architectures present an alternative geometry that can simplify fabrication and may increase compatibility with flexible substrates \cite{Planar_Intro,planar_intro2,planar_intro3}. In the planar configuration, the two electrodes are patterned laterally on the same substrate surface, with the active-layer bridging the gap between them. In this geometry, the voltage is applied to one of the electrodes to establish an electric field across the lateral channel, and resistive switching occurs within the region of the active-layer between the electrodes. Unlike charge-based memories, RRAM does not rely on charge storage but instead on the formation and rupture of conductive filaments (CFs) \cite{filament, filament1, filament2}.

In filamentary switching, CFs originate from the field-driven migration of oxygen ions or metal cations within the active-layer, producing localized conductive pathways that can be repeatedly formed and ruptured \cite{oxygen_vac,fil_phys}. When a metal oxide serves as the active-layer, a positive bias applied to the TE drives oxygen anions toward the electrode, leaving behind a chain of oxygen vacancies that coalesce into a CF \cite{oxygen_vac2}. This oxygen-deficient region places the device in a low-resistive state (LRS) \cite{States}. A subsequent negative bias reverses this process, driving oxygen anions back into the filament region, where recombination with vacancies ruptures the CF and restores insulating behavior, thereby switching the device into a high-resistive state (HRS) \cite{states1, states2, states3}. The initial creation of a CF is referred to as the forming process, while subsequent rupture and reformation correspond to RESET and SET processes, respectively. To prevent permanent breakdown and to control filament dimensions, a compliance current limit (CCL) is applied during forming and SET operations \cite{CCL1}. The CCL directly determines filament diameter and also the resulting LRS value \cite{CCL}.

While RRAM has been extensively investigated across various materials and device geometries, comparatively few studies have examined how active-layer architecture influences switching power and memory window characteristics in planar RRAM devices. In this work, planar ITO/WO$_x$/ITO RRAM structures were fabricated with either compact TF active-layers or porous helical active-layers deposited by glancing angle deposition (GLAD). The helical morphology was selected because its porous, interconnected geometry increases surface area relative to TF layers and provides reproducible, tunable parameters for systematic study \cite{GLAD}.

Accordingly, this study aims to (i) determine whether planar RRAM devices exhibit dependence on active-layer thickness across the $50$--$200$ nm range, (ii) evaluate the ability of porous helical architectures to sustain operation at reduced CCLs, and (iii) investigate how current scaling influences the memory window in planar devices. Through this comparison of TF and helical architectures, the work seeks to clarify how geometry, rather than thickness, governs the low-power and memory characteristics of planar RRAM, and to assess the potential of helical designs for enabling energy-efficient and flexible memory technologies. 

\section{Materials and Methods}
In this study, two devices with differing active-layer architectures were investigated. The first active-layer architecture was a traditional TF layer, while the second active-layer utilized a helical geometry fabricated by a GLAD method. Both device types were fabricated with active-layer thicknesses of $50$, $100$, and $200$~nm to confirm that the observed switching behavior was not dominated by thickness effects. Both device architectures used indium tin oxide (ITO) as electrodes with a tungsten oxide (WO$_x$) active-layer. The ITO/WO$_x$/ITO metal-insulator-metal (MIM) planar configuration was fabricated on a manufactured ITO-coated glass slide substrate (SPI Supplies). While devices in this study were tested in a planar configuration, an ITO-coated glass slide substrate was chosen to facilitate potential future vertical studies and transparent device testing. 

Before deposition, the substrates were cleaned with deionized (DI) water and acetone, followed by an oxygen plasma treatment in a Harrik Plasma PDC-$32C$ system. Fabrication of the active-layer and top electrode layer utilized tungsten trioxide (WO$_3$) pellets and ITO (In$_2$O$_3$/SnO$_2$) pellets (Kurt J. Lesker, $99.99\%$) as source materials, which were evaporated via physical vapor deposition (PVD) using electron-beam (e-beam) heating without further purification. During e-beam evaporation of WO$_3$, partial reduction to a sub-stoichiometric state (WO$_{3-x}$) can occur \cite{Reduction,Reduction1,Sabrina}. Because all active-layers were deposited under identical conditions, any deviation from stoichiometry was consistent across samples. To reflect this uncertainty, the deposited material is referred to as WO$_x$ throughout this work. Deposition rates were measured using an Ificon SQC-$310$C deposition controller with a quartz-crystal microbalance in vacuo, and all depositions were conducted at a pressure of $10^{-6}$~torr. 

To fabricate the TF devices, the ITO substrate was attached to the substrate holder of the PVD system with double-sided carbon tape on the glass side of the substrate with no inclination. The WO$_x$ active-layer was then deposited at an average rate of $0.2$~nm/s until the desired thickness of $50$, $100$, or $200$~nm was achieved. Once the active-layer was deposited, a shadow mask was applied to pattern circular ITO electrodes ($50$~$\mu$m diameter, $150$~nm thick), deposited at an average rate of $0.3$~nm/s. The TF devices in this study were used as-deposited, with no further post-deposition treatments. The as-deposited ITO electrodes exhibited a noticeable color shift, consistent with the amorphous state typically observed in ITO before annealing. This color change arises because the lack of long-range crystallinity alters the optical band structure and light scattering compared to crystalline ITO \cite{ITO_Color, ITO_Color2}.

\begin{figure}[H]
    \centering
    \includegraphics[width=.65\textwidth]{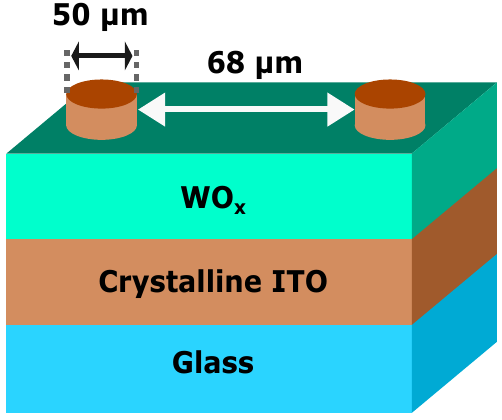}
    \caption{Schematic cross-section of the TF device stack. Vertical thicknesses are not drawn to scale, and the schematic is intended only to illustrate the layer sequence and lateral electrode geometry.}
    \label{fig:TF_Struct}
\end{figure}

\begin{figure}[H]
    \centering
    \includegraphics[width=.6\textwidth]{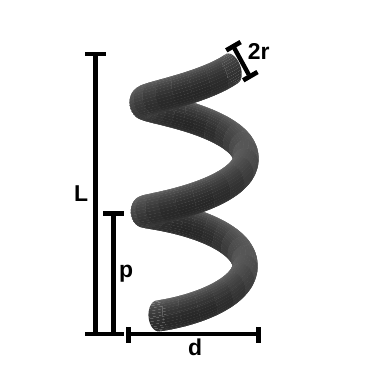}
    \caption{Model of a helix with parameters: length ($L$), pitch ($p$), helix diameter ($d$), and rod thickness ($2r$).}
    \label{fig:helix}
\end{figure}

Figure \ref{fig:TF_Struct} depicts a schematic of the TF device architecture, including the ITO-coated glass substrate, WO$_x$ active-layer, and patterned ITO electrodes. The lateral electrode geometry was measured with a Keyence VHX-$2000$ digital microscope. As mentioned previously, the helical devices were fabricated using a GLAD method. In this technique, the substrate is positioned at an angle ($\alpha$), typically greater than $70^\circ$, such that the vapor flux impinges on the substrate at an oblique incidence. Film growth is governed by atomic self-shadowing and limited surface diffusion, leading to highly porous and anisotropic nanostructures \cite{Helix_Parameters}. By controlling $\alpha$ and dynamically manipulating the substrate, GLAD enables fabrication of architectures such as slanted columns, chevrons, and, in this work, helices \cite{Chevron,nanorods}. Figure \ref{fig:helix} depicts a model of a helix with parameters controlled by deposition rate, $\alpha$, and substrate rotation speed.

The geometric parameters of the helices in Fig.~\ref{fig:helix} are defined as follows. The pitch ($p$) is the vertical distance along the helix axis corresponding to one complete $360^\circ$ rotation. The diameter ($d$) is the lateral width of the coil, derived from $\alpha$ and the shadowing effect. The rod radius ($r$) is half the rod thickness, with the rod thickness reported as $2r$ determined by the column growth rate. The helix length ($L$) corresponds to the total vertical extent of the deposited helix, similar to the thickness of a TF layer, determined by deposition time and rate. The pitch can be approximated as the ratio of deposition rate to rotation rate, as expressed in Equation~\ref{eq:Helix} \cite{Helix}:

\begin{equation}
    p \approx \frac{R_d}{f},
    \label{eq:Helix}
\end{equation}
where $p$ is the pitch length in nm, $R_d$ is the deposition rate in nm/s, and $f$ is the substrate rotation rate in revolutions/s.
 
To deposit the helical active-layer, the ITO substrate was attached to the PVD substrate holder with double-sided tape and inclined to $\mathord{\mathord{\sim}}86^\circ$. The parameter $\alpha \approx 86^\circ$ was chosen because it produces separated nano-columns in this material system. With $p$ fixed to $50$~nm via Eq.~\ref{eq:Helix}, a feedback loop, controlled by a software connected to the machine, adjusted $f$ for fluctuations in $R_d$. The average substrate rotation rate was $0.238$~rpm for a deposition rate of $0.2$~nm/s.

\begin{figure}[H]
    \centering
    \includegraphics[width=.75\textwidth]{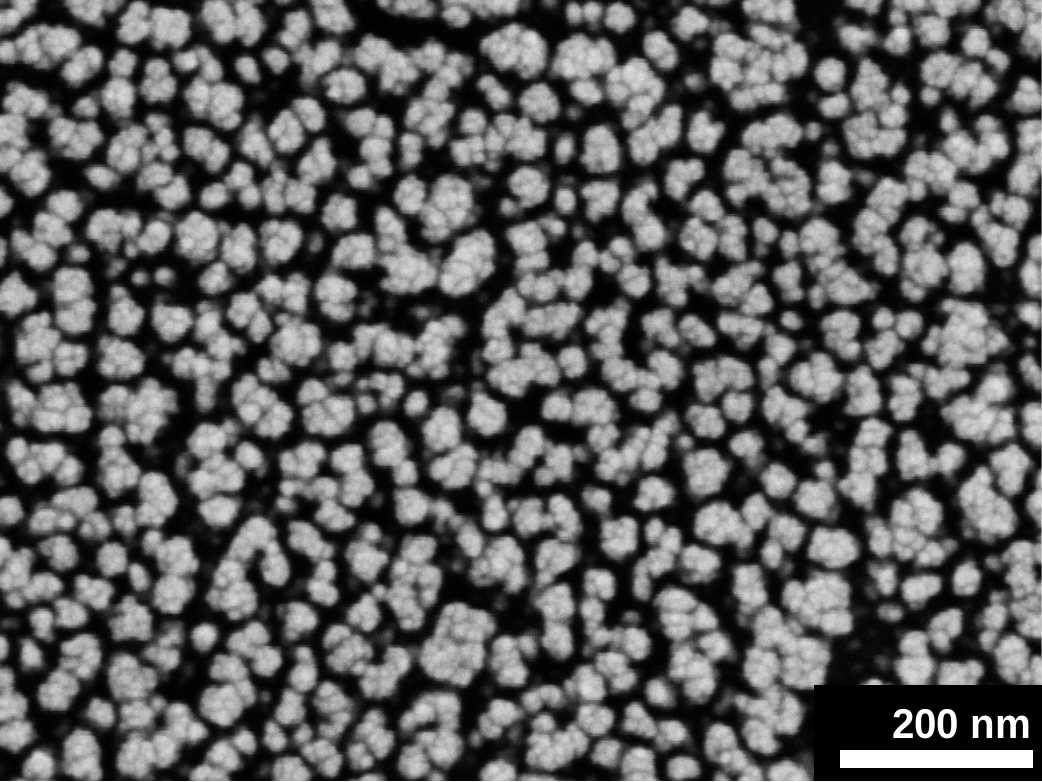}
    \caption{Top-view SEM image of the porous helical WO$_x$ active-layer.}
    \label{fig:helix_SEM}
\end{figure}

Similarly to the TF devices, helical devices with active-layer lengths of $50$, $100$, and $200$~nm were fabricated. Finally, the ITO planar electrodes were deposited without inclination. This $150$~nm layer, patterned into $50$~$\mu$m diameter circles by a shadow mask, was grown at an average rate of $0.3$~nm/s. The lateral electrode geometry was again measured with a Keyence VHX-$2000$ digital microscope. A Zeiss Supra $40$VP field-emission scanning electron microscope (SEM) was used to capture Fig.~\ref{fig:helix_SEM}, which depicts a top-view SEM image of the porous helical active-layer of the $50$~nm devices. The circular lateral features are consistent with arrays of helices \cite{Top_view}.

\begin{figure}[H]
    \centering
    \includegraphics[width=.75\textwidth]{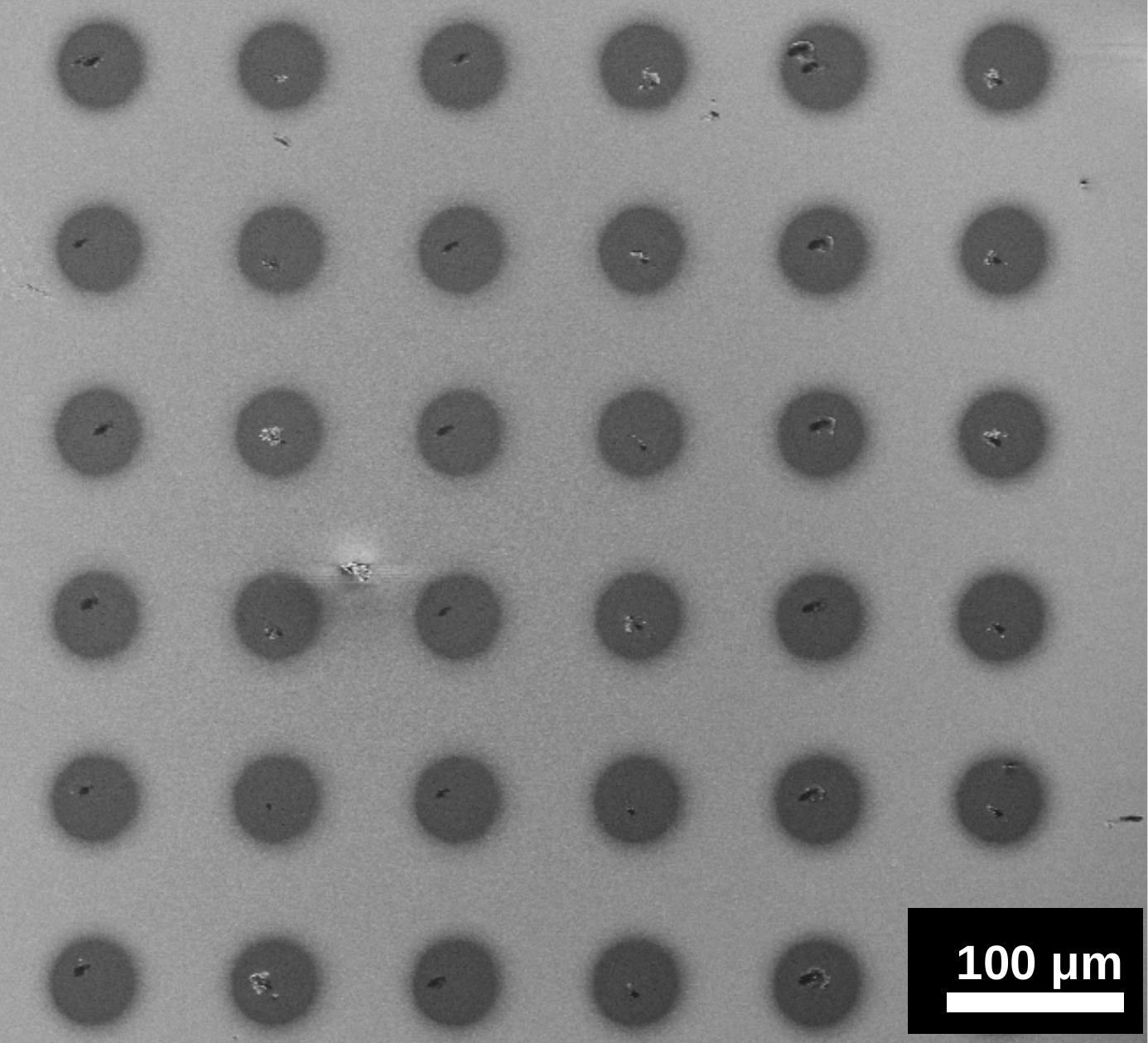}
    \caption{SEM top-view of porous helical WO$_x$ devices.}
    \label{fig:devices}
\end{figure}

Figure \ref{fig:devices} shows a section of helical devices post-testing. The circular regions correspond to the patterned ITO electrodes, while the surrounding matrix is the porous WO$_x$ active-layer. The image shows well-defined electrode edges and uniform separation between pads. Marks on the pads indicate probing locations during electrical testing.

Electrical characterization was performed using an EPS-$150$ probe station equipped with tungsten probe tips ($\sim$$5$~$\mu$m tip radius) and a Keysight B$1500$ semiconductor device analyzer. The RRAM devices were characterized in a planar configuration between patterned ITO electrodes. This geometry provides well-defined contacts and reduces complications from vertical injection barriers due to the symmetric electrode configuration \cite{vertical_injection_barrier}. For compact TF devices, planar testing allows intrinsic switching to be studied without stack-related artifacts \cite{Planar_testing}. For porous helical devices, it directly probes conduction through the interconnected network and minimizes the risk of premature shorts. In both cases, planar measurements establish a baseline understanding of resistive switching.

In all experiments, the right planar electrode was grounded, and a bias was applied to the left electrode. Both TF and helical devices operated with a CCL of $5$~mA, but only helical devices could function at $500$~$\mu$A. TF devices did not sustain operation at $500$~$\mu$A. For TF and helical devices at $5$~mA, the operating parameters were identical. Devices were initially formed from the as-deposited state with a $7$~V bias. To RESET, a $-3$~V bias was applied, while SET was performed with a $2$~V bias. For helical devices at $500$~$\mu$A, the forming voltage remained $7$~V, but RESET and SET biases decreased to $-1.5$~V and $1.5$~V, respectively. 

Each device was cycled $50$ times to suppress first-cycle transients and capture stable switching statistics. Extended endurance testing was outside the scope of this work, which primarily focused on low-power operation. A device was classified as successful if it underwent proper electroforming and subsequently exhibited reproducible switching between HRS and LRS for $50$ consecutive cycles. 

A total of $180$ successful devices are reported. The operation yields were $74.07$\% ($60/81$) for TF devices at $5$~mA, $81.08$\% ($60/74$) for helical devices at $5$~mA, and $85.71$\% ($60/70$) for helical devices at $500$~$\mu$A. The $60$ successful devices of each category included $20$ devices at each oxide thickness ($50$, $100$, $200$~nm). Since all thicknesses shared the same planar structure and operating parameters, they are considered a single device type for analysis. Results are nevertheless presented by thickness to confirm that no systematic thickness dependence was observed. For statistical comparisons, the per-device median over $50$ cycles was used to represent stable switching behavior.   

\begin{figure}[H]
    \centering
    \includegraphics[width=1\textwidth]{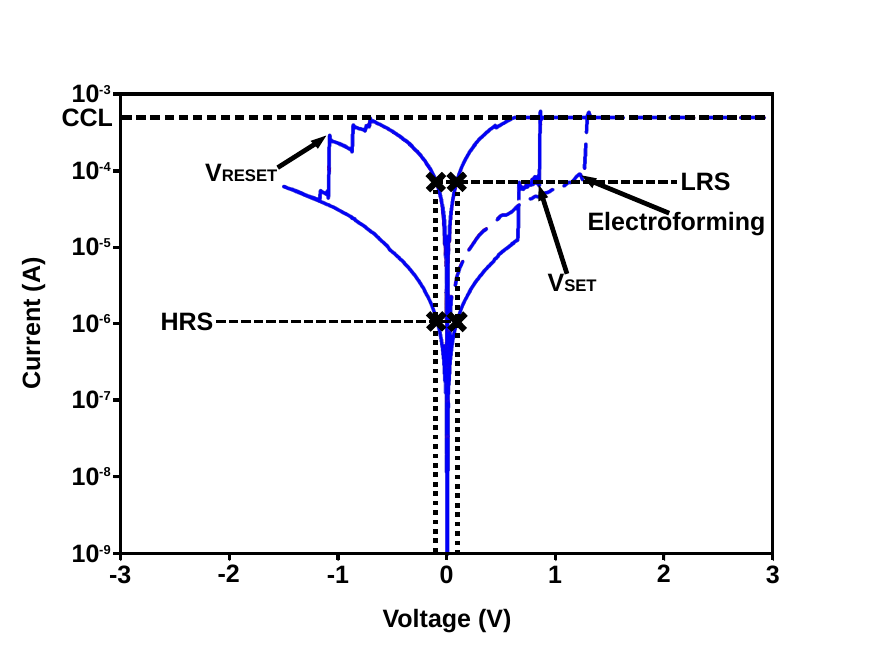}
    \caption{Bipolar I-V characteristic of a typical $500$~$\mu$A CCL helical device cycle. Electroforming, SET, and RESET processes are indicated, with corresponding $V_{\text{SET}}$ and $V_{\text{RESET}}$. HRS and LRS are indicated by $\times$ markers at a non-disturbing read bias of $0.1$~V.}
    \label{fig:cycle}
\end{figure}

Figure \ref{fig:cycle} shows a representative bipolar I-V curve from the median cycle of a randomly selected successful device, illustrating the key switching features analyzed in this work: $V_{\text{SET}}$, $V_{\text{RESET}}$, and the corresponding HRS and LRS states. $V_{\text{SET}}$ was defined as the bias at which the device switched from HRS to LRS during the positive sweep, identified by the abrupt current increase reaching the CCL. $V_{\text{RESET}}$ was defined as the bias at which the device switched from LRS to HRS during the negative sweep, identified by the sudden current drop. Small current steps observed before these abrupt transitions were attributed to defect-assisted conduction or partial filament formation and were not considered true SET/RESET events since they did not produce stable state changes. The switching currents ($I_{\text{SET}}$ and $I_{\text{RESET}}$) were taken at $V_{\text{SET}}$ and $V_{\text{RESET}}$, respectively. Using Equation~\ref{eq:power}:

\begin{equation}
    P = IV,
    \label{eq:power}
\end{equation}
the instantaneous switching power was derived for both SET and RESET switching nodes, where $P$ is instantaneous power, $I$ is switching current, and $V$ is applied voltage, the instantaneous switching power was quantified. HRS and LRS states were measured at $0.1$~V after cycling. This read voltage was chosen as it is sufficiently low to avoid disturbing the state while still yielding measurable current. The memory window, corresponding to the separation between HRS and LRS currents in Fig.~\ref{fig:cycle}, was defined as the ratio of $I_{\text{LRS}}$ to $I_{\text{HRS}}$ at $0.1$~V, as given in Equation~\ref{eq:MW} \cite{Mem_window_eq}:

\begin{equation}
    MW = \frac{I_{\text{LRS}}}{I_{\text{HRS}}}.
    \label{eq:MW}
\end{equation}
where  $MW$ is the memory window, $I_{\text{LRS}}$ is the current obtained at the LRS, and $I_{\text{HRS}}$ is the current obtained at the HRS. This definition was used to extract the memory window from each device for subsequent statistical analysis.

\section{Results and Discussion}
The electrical switching characteristics of planar devices with different active-layer architectures were investigated to evaluate low-power operation and reproducibility. For TF devices, successful operation was only observed at a CCL of $5$~mA. When biased below this value, TF devices typically cycled only $2$--$3$ times before failing to produce a distinguishable memory window. In contrast, helical devices remained functional at a significantly lower CCL of $500$~$\mu$A. This behavior is attributed to the porous helical geometry, which enhances local electric fields and confines the effective switching volume \cite{controlled_CF}. Such field concentration promotes controlled filament nucleation and stabilization at reduced current levels. By comparison, TF devices lack this geometric field enhancement and therefore require higher compliance currents and voltages to achieve stable filament formation and reproducible switching.

\begin{figure}[H]
    \centering
    \includegraphics[width=.75\textwidth]{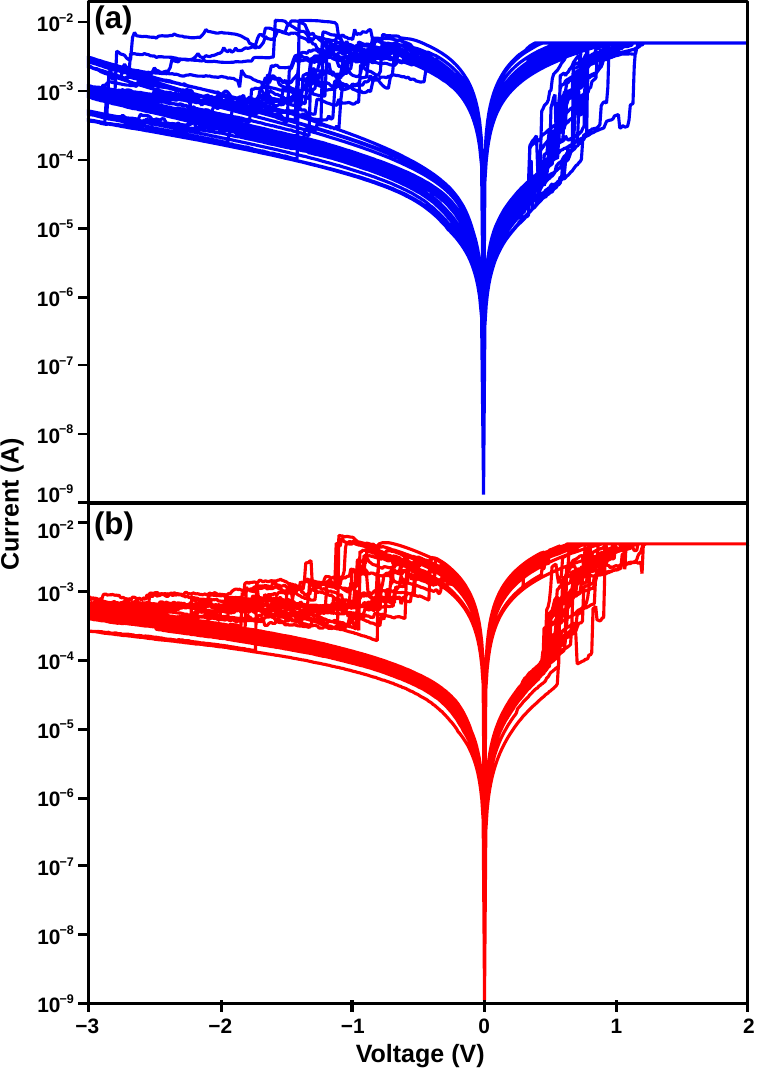}
    \caption{Twenty consecutive bipolar I--V cycles of (a) helical and (b) TF devices at $5$~mA CCL.}
    \label{fig:cycles}
\end{figure}

Figure \ref{fig:cycles} compares consecutive cycling of helical (a) and TF (b) devices operating at $5$~mA CCL. Both exhibit reproducible bipolar switching with broadly similar cycle shapes, making it difficult to draw firm conclusions from visual inspection alone. Therefore, statistical analysis of key switching parameters was performed to quantitatively assess differences between device types and operating conditions. Full cycling overlays at $500$~$\mu$A are not shown, as their shapes are qualitatively identical to those at $5$~mA.

\begin{figure}[H]
    \centering
    \includegraphics[width=1\textwidth]{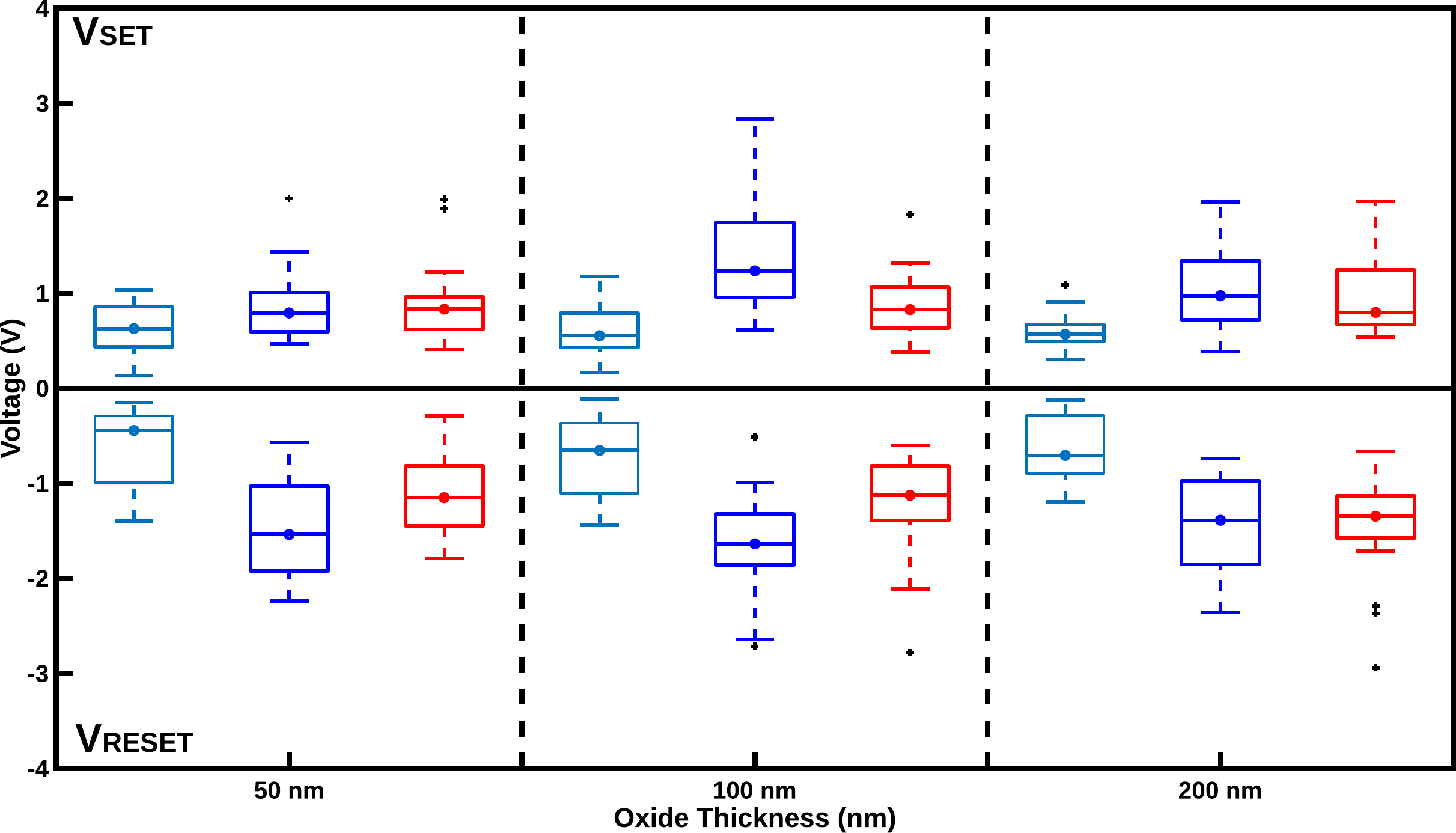}
    \caption{Statistical distribution of SET/RESET voltages across oxide thicknesses of $50$, $100$, and $200$~nm. Left: helical devices at $500$~$\mu$A CCL; middle: helical devices at $5$~mA CCL; right: TF devices at $5$~mA CCL.}
    \label{fig:voltage}
\end{figure}

Figure \ref{fig:voltage} summarizes the statistical distributions of SET and RESET voltages as a function of oxide thickness. No systematic dependence on thickness is observed, indicating that switching thresholds are largely geometry-independent across the $50$--$200$~nm range. At $5$~mA CCL, TF and helical devices exhibit comparable $V_{\text{SET}}$ and $V_{\text{RESET}}$ values, confirming that both device types switch under similar electrical conditions at high current. In contrast, helical devices operating at $500$~$\mu$A show significantly reduced $V_{\text{RESET}}$ values, decreased by $\sim$$60$\% relative to helical devices at $5$~mA, while maintaining similar $V_{\text{SET}}$ values. This reduction in RESET bias reflects easier rupture of the CF at lower CCL, consistent with more controlled filament growth and stabilization in the helical geometry \cite{low_power_rupture}.

\begin{figure}[H]
    \centering
    \includegraphics[width=1\textwidth]{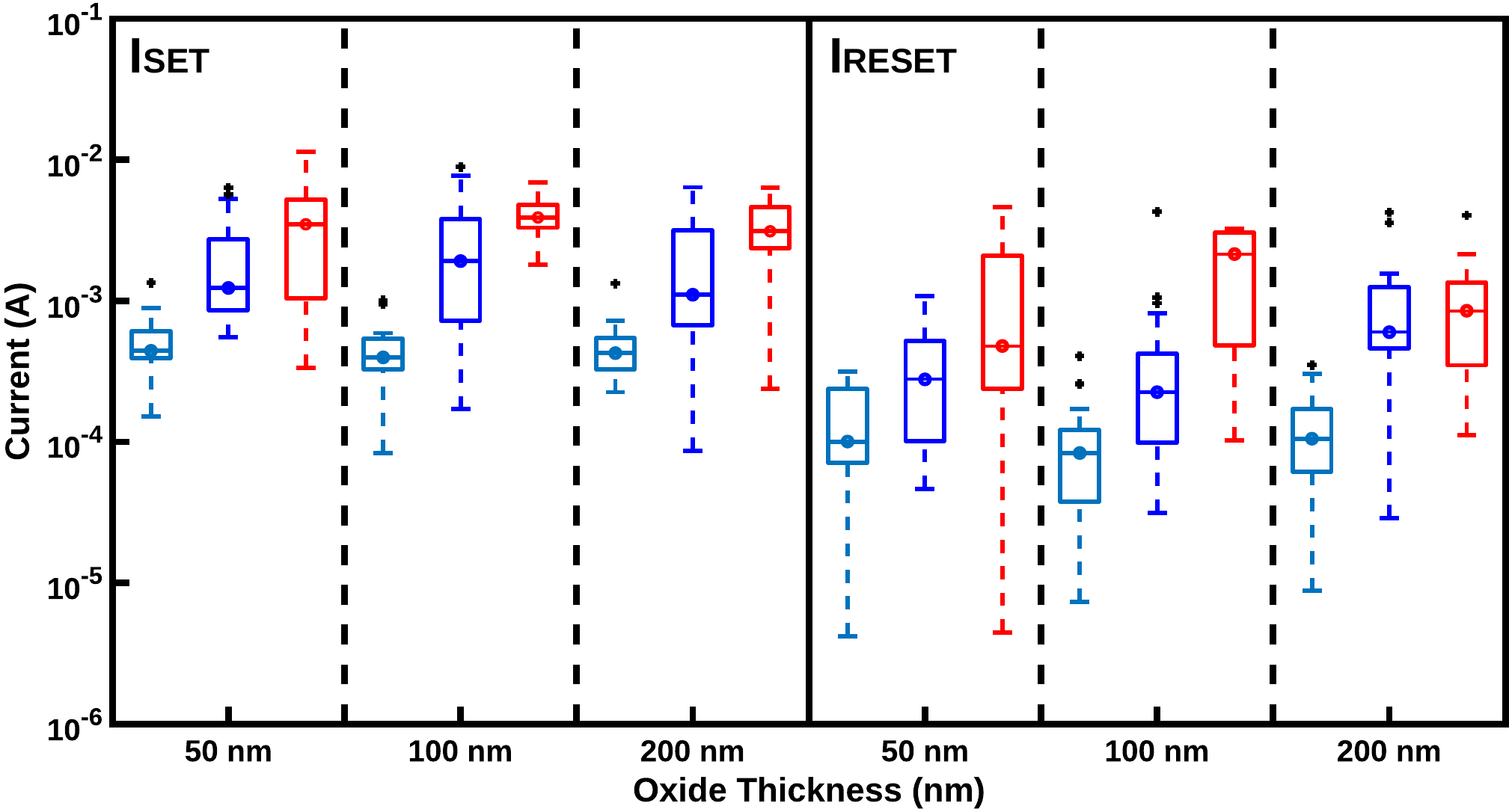}
    \caption{Statistical distribution of SET/RESET currents across oxide thicknesses of $50$, $100$, and $200$~nm. Left: helical devices at $500$~$\mu$A CCL; middle: helical devices at $5$~mA CCL; right: TF devices at $5$~mA CCL.}
    \label{fig:current}
\end{figure}

Figure \ref{fig:current} shows the distribution of switching currents as a function of oxide thickness. In all cases, $I_{\text{SET}}$ exceeds $I_{\text{RESET}}$, reflecting the higher current required to form a CF compared to rupturing it. At $5$~mA CCL, TF and helical devices exhibit similar switching currents, with helices having slightly lower average current, although the distributions overlap. In contrast, helical devices operating at $500$~$\mu$A consistently exhibit the lowest $I_{\text{SET}}$ and $I_{\text{RESET}}$ values, with average reductions of $\sim$$68$\% and $\sim$$75$\%, respectively, compared to their $5$~mA counterparts. This confirms that stable switching can be sustained at substantially reduced current.

\begin{figure}[H]
    \centering
    \includegraphics[width=1\textwidth]{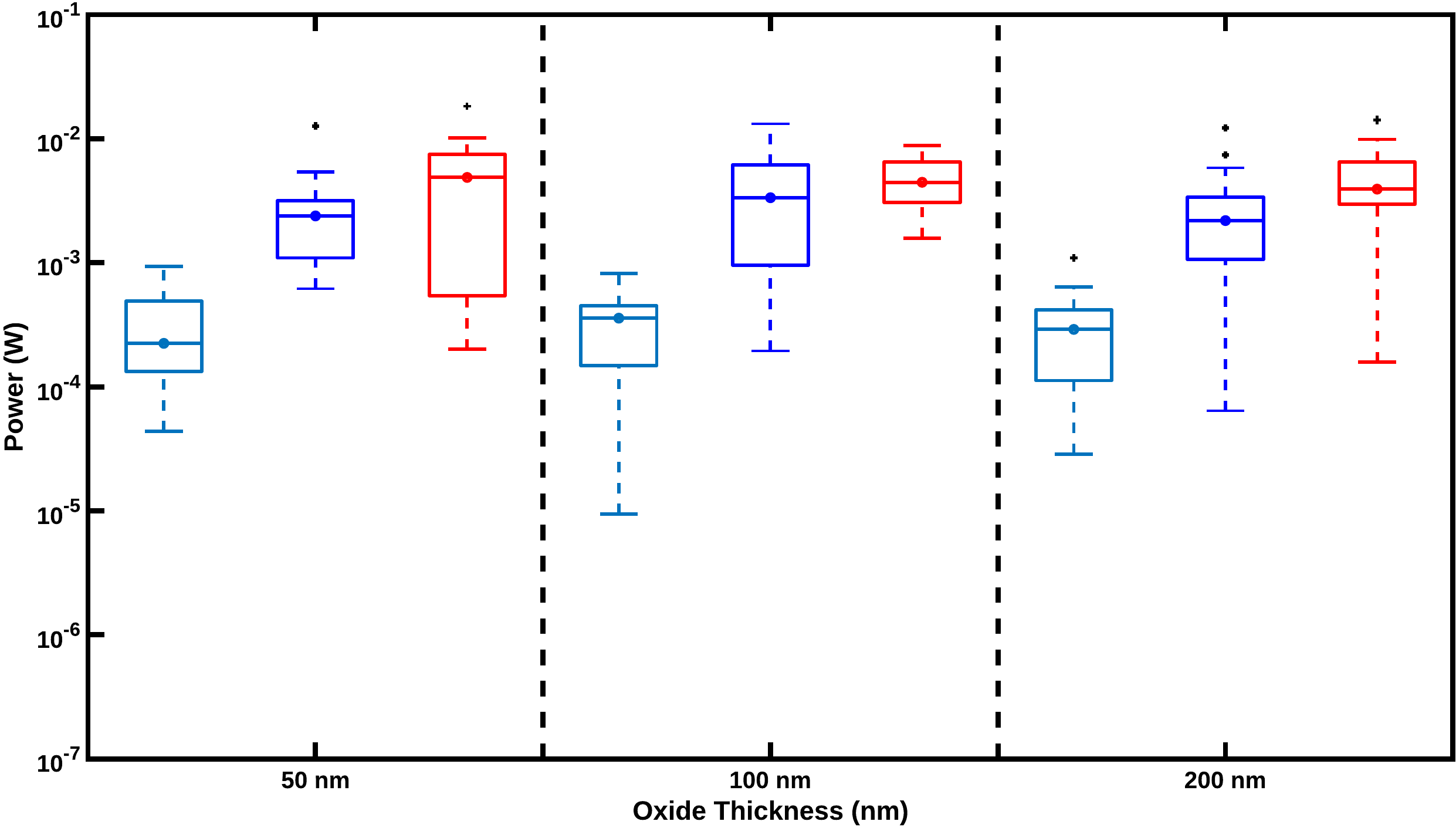}
    \caption{Statistical distribution of SET power across oxide thicknesses of $50$, $100$, and $200$~nm. Left: helical devices at $500$~$\mu$A CCL; middle: helical devices at $5$~mA CCL; right: TF devices at $5$~mA CCL.}
    \label{fig:set_power}
\end{figure}

\begin{figure}[H]
    \centering
    \includegraphics[width=1\textwidth]{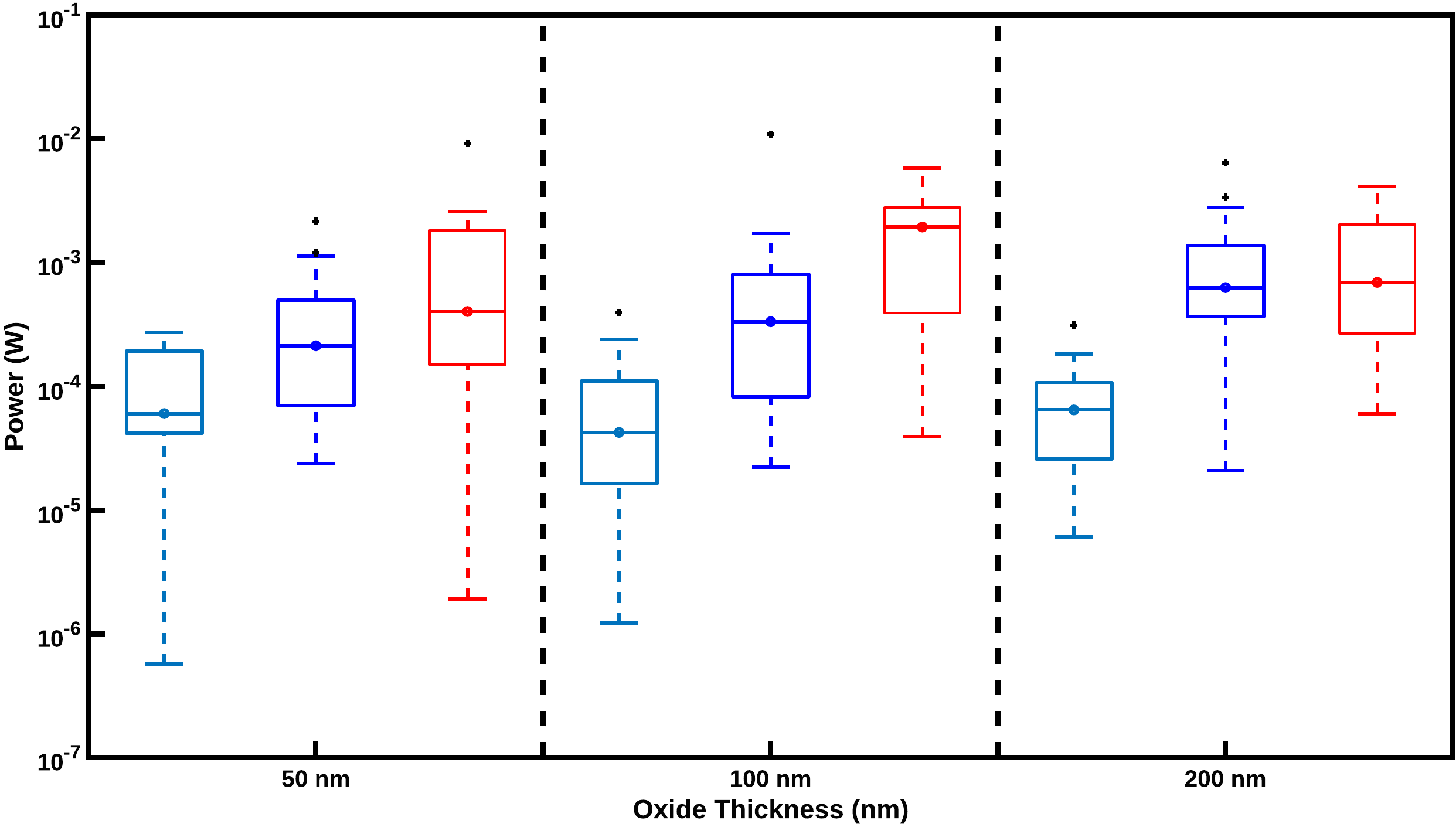}
    \caption{Statistical distribution of RESET power across oxide thicknesses of $50$, $100$, and $200$~nm. Left: helical devices at $500$~$\mu$A CCL; middle: helical devices at $5$~mA CCL; right: TF devices at $5$~mA CCL.}
    \label{fig:reset_power}
\end{figure}

Figures \ref{fig:set_power} and \ref{fig:reset_power} show the distributions of instantaneous SET and RESET switching power, respectively, calculated from Equation~\ref{eq:power}. In all cases, RESET power is lower than SET power, consistent with the reduced energy required to rupture CFs compared to forming them. At $5$~mA CCL, TF and helical devices again exhibit broadly similar power requirements, with overlapping distributions. In contrast, helical devices operating at $500$~$\mu$A show marked reductions, with average decreases of $\sim$$89$\% in SET power and $\sim$$83$\% in RESET power compared to their $5$~mA counterparts. These results highlight the advantage of the helical architecture in sustaining reproducible switching at substantially reduced power levels, a regime inaccessible to TF devices.

\begin{figure}[H]
    \hspace{-1.4cm}
    \includegraphics[width=1.2\textwidth]{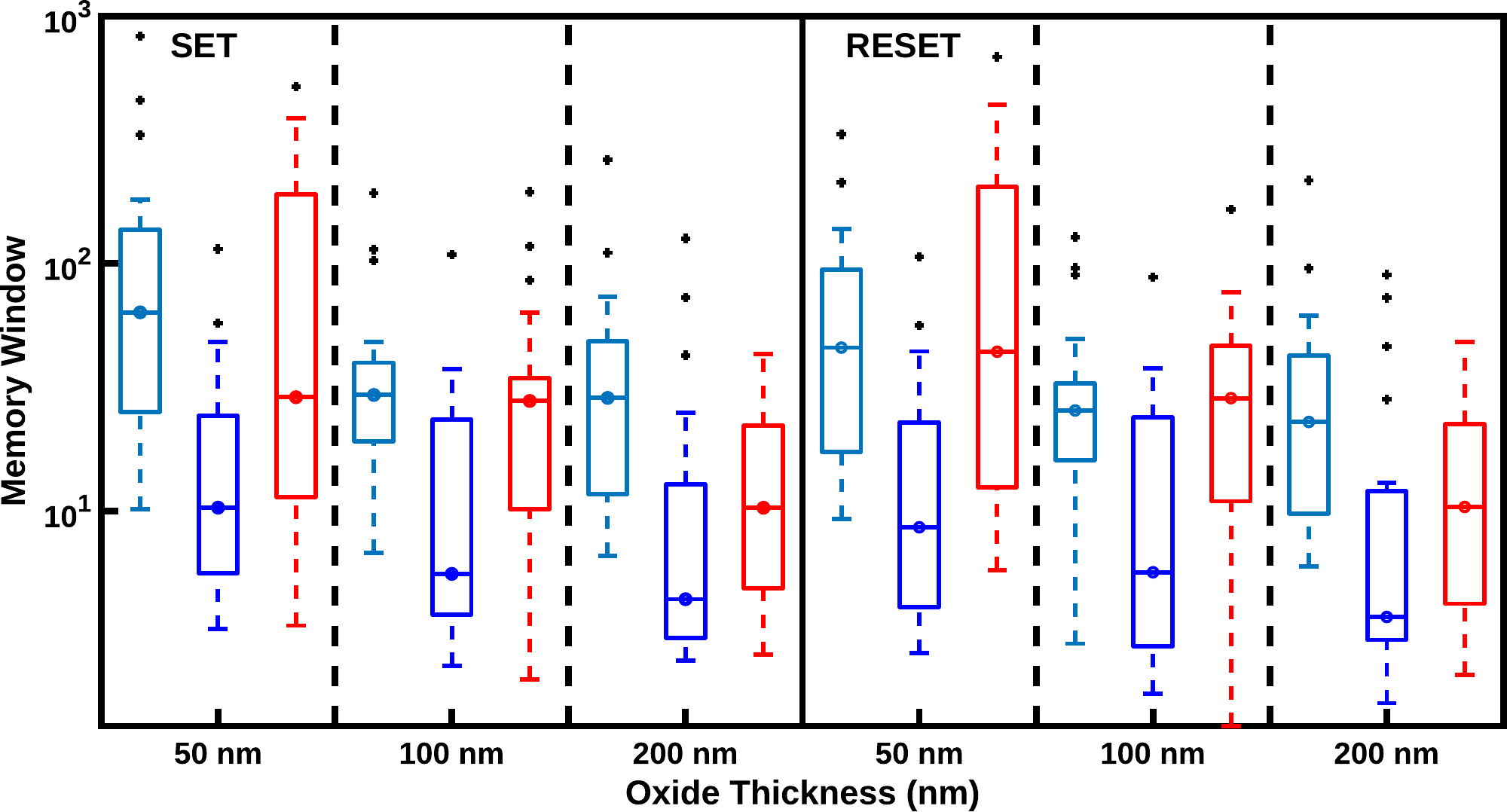}
    \caption{Statistical distribution of LRS and HRS currents across oxide thicknesses of $50$, $100$, and $200$~nm. Left: helical devices at $500$~$\mu$A CCL; middle: helical devices at $5$~mA CCL; right: TF devices at $5$~mA CCL.}
    \label{fig:HRS_LRS}
\end{figure}

\begin{figure}[H]
    \centering
    \includegraphics[width=1\textwidth]{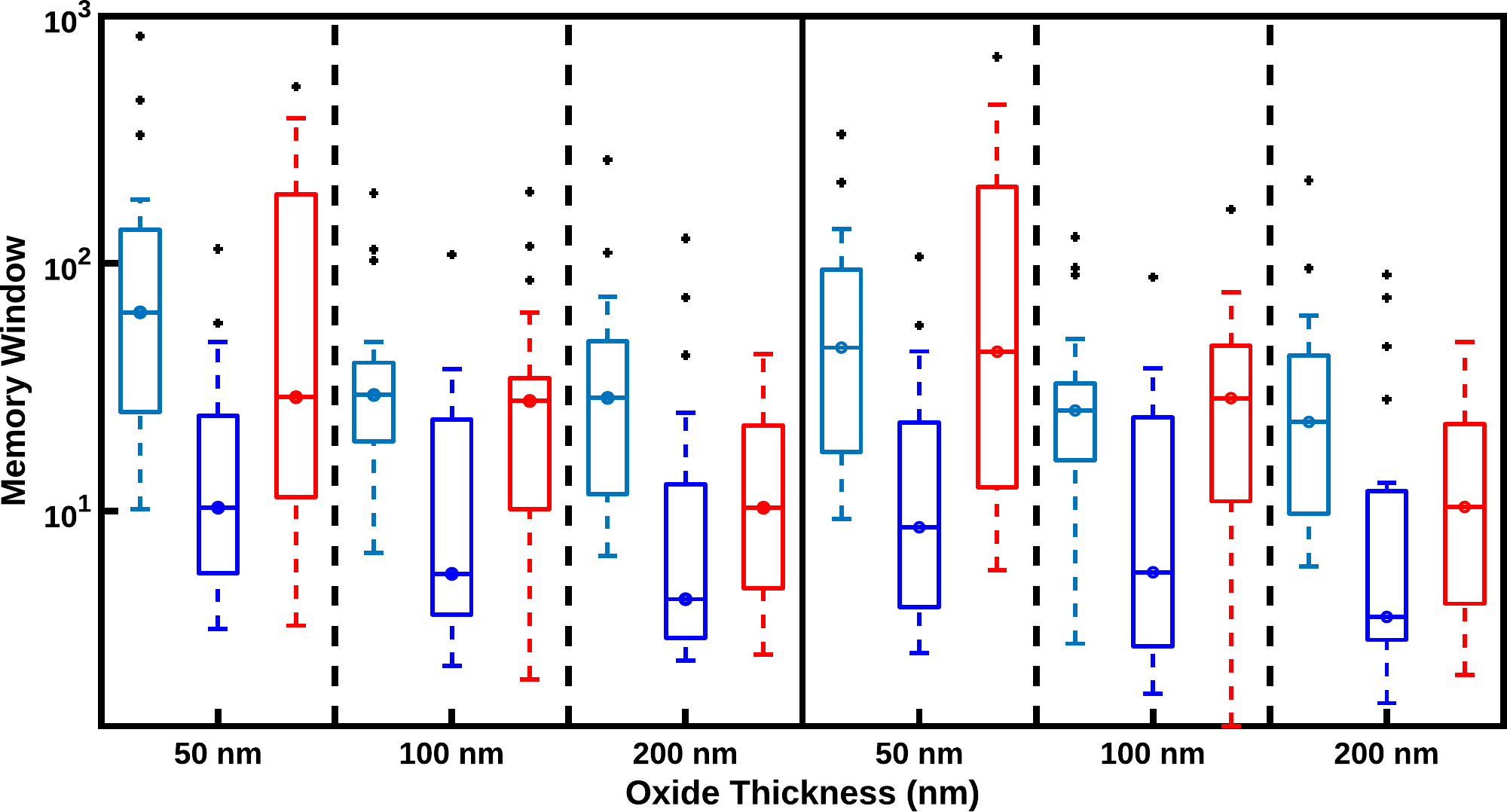}
    \caption{Statistical distribution of memory windows across oxide thicknesses of $50$, $100$, and $200$~nm. Left: helical devices at $500$~$\mu$A CCL; middle: helical devices at $5$~mA CCL; right: TF devices at $5$~mA CCL.}
    \label{fig:Mem_Win}
\end{figure}

The distributions in Figure~\ref{fig:HRS_LRS} show LRS and HRS currents, while Figure~\ref{fig:Mem_Win} summarizes the corresponding memory windows derived from Equation~\ref{eq:MW}. No systematic thickness dependence is evident. Across all device types, LRS currents exceed HRS currents by one to two orders of magnitude, establishing a clear resistive switching window. At $5$~mA CCL, TF and helical devices exhibit similar LRS and HRS values, with helical devices trending slightly lower on average but with overlapping distributions. Helical devices at $500$~$\mu$A maintain similar LRS currents, reduced by only $\sim$$17$\% relative to $5$~mA operation, but show a pronounced reduction in HRS current, decreased by $\sim$$81$\%. Although both states are suppressed, the stronger reduction in HRS directly enlarges the memory window.

As shown in Figure~\ref{fig:Mem_Win}, the memory window for helical devices at $5$~mA CCL is modest, with a median of $\sim$$7$ across thicknesses, whereas TF devices average $\sim$$22$ under the same conditions. When the CCL is reduced to $500$~$\mu$A, the memory window of helical devices expands dramatically, averaging $\sim$$44$ across thicknesses. This represents a $5$--$7$$\times$ improvement relative to helical devices at $5$~mA. These results confirm that reducing the CCL in helical devices not only lowers power consumption but also substantially enhances the read margin between HRS and LRS. The lower compliance selectively suppresses leakage in the HRS more strongly than it limits conduction in the LRS, yielding a more robust memory window \cite{HRS_leakage}. This effect is inaccessible to TF devices, which fail to sustain reproducible switching below $5$~mA. Together, the combined current and memory window analysis highlights the advantage of the helical geometry in enabling both low-power operation and improved state separation, two key metrics for viable resistive switching memory.

\begin{figure}[H]
    \centering
    \includegraphics[width=1\textwidth]{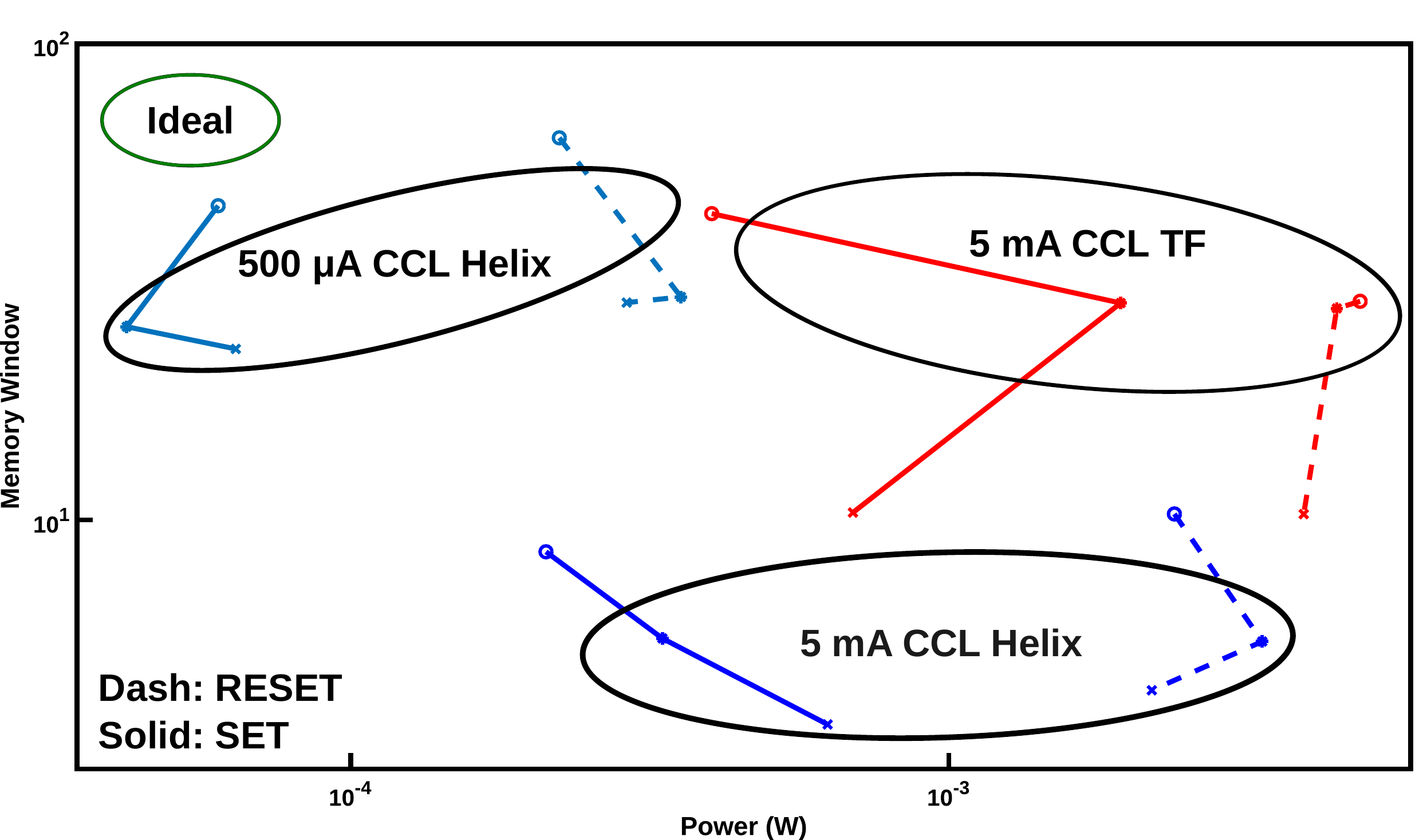}
    \caption{Switching power versus memory window for TF ($5$~mA) and helical devices ($5$~mA, $500$~$\mu$A). Symbols: $\circ = 50$~nm, $* = 100$~nm, $\times = 200$~nm. Helical devices at $500$~$\mu$A achieve the lowest power and largest memory windows, closest to the ideal regime.}
    \label{fig:MemvPow}
\end{figure}

Figure \ref{fig:MemvPow} illustrates the trade-off between switching power and memory window. TF devices and helical devices at $5$~mA cluster in a regime of relatively high power and modest memory windows, while only helical devices at $500$~$\mu$A approach the ideal combination of low switching power and enlarged read margin. The absence of any systematic dependence on oxide thickness underscores that this benefit arises from geometry, specifically the porous helical morphology of the active-layer. Together, these results establish that the helical architecture enables a fundamentally different scaling pathway for resistive switching, where low-power operation and robust memory windows are simultaneously realized, in contrast to the limitations of conventional TF devices.

\section{Conclusions}
In this study, planar ITO/WO$_x$/ITO RRAM devices incorporating either compact thin-film (TF) or porous helical active-layers fabricated by glancing angle deposition (GLAD) were systematically compared. Across three nominal oxide thicknesses ($50$, $100$, and $200$ nm) and $180$ functional devices, no systematic dependence of switching behavior on thickness was observed. This result indicates that planar RRAM devices are intrinsically robust to variations in active-layer thickness, an advantage for flexible and large-area electronics where thickness non-uniformity may arise from strain, curvature, or deposition variability.

At a compliance current limit (CCL) of $5$~mA, TF and helical devices exhibited comparable switching thresholds and power, with SET currents higher than RESET currents as expected, and RESET power lower than SET power. However, TF devices failed to sustain reproducible operation below $5$~mA, whereas helical devices remained functional at $500$~$\mu$A. Reducing the CCL from $5$~mA to $500$~$\mu$A in helical devices lowered the RESET voltage by $\sim$$60$\% while maintaining similar SET voltages. Switching currents decreased by $\sim$$68$\% (SET) and $\sim$$75$\% (RESET), yielding corresponding power reductions of $\sim$$89$\% and $\sim$$83$\%, respectively.

Operation at reduced CCL was also shown to enlarged the memory window. Although both LRS and HRS currents decreased, the reduction was asymmetric: the LRS current dropped by $\sim$$17$\%, whereas the HRS current decreased by $\sim$$81$\%. This asymmetry increased the median memory window from $\sim$$7$ at $5$~mA to $\sim$$44$ at $500$~$\mu$A, representing a $5$--$7$$\times$ improvement. When plotted against switching power, helical devices at $500$~$\mu$A occupied the desirable regime of lowest power and largest memory window compared to both TF devices and higher-current helical devices.

These results support a model in which the porous helical morphology concentrates local electric fields and confines the effective switching volume, enabling controlled filament nucleation and rupture at reduced current. Under such conditions, HRS leakage is selectively suppressed relative to LRS conduction, thereby enlarging the memory window while lowering power consumption. In contrast, compact TF layers lack comparable field localization and volumetric confinement, requiring higher currents to stabilize filaments and failing to sustain operation below the milliampere level.

Several boundaries of this study define opportunities for future work. First, several characteristics: endurance, retention, and variability under pulsed operation were not addressed here and should be investigated to establish long-term reliability. Second, while the planar geometry isolates intrinsic switching, extending the helical active-layer architecture to vertical stacks and dense arrays will be critical for application relevance. Finally, the demonstrated role of geometry and surface area suggests exploration of alternative high-surface-area materials, such as hydrogels or aerogels \cite{Hydrogel_Aerogel_Porous}, which may further enhance field concentration and filament confinement. Coupled with the planar ITO/WO$_x$ platform, these approaches point toward transparent, flexible, and energy-efficient RRAM devices.

\section{Acknowledgments}
This material is based upon work supported by the National Science Foundation under Grant No. 2425226.

\section{Conflict of Interest}
The authors declare that they have no conflict of interest.
\newpage
\singlespacing

\bibliography{main} 
\bibliographystyle{unsrt}

\end{document}